\documentclass[aps,preprint]{revtex4}%
\usepackage{amsfonts}
\usepackage{amsmath}
\usepackage{amssymb}
\usepackage{graphicx}%
\begin{document}
\preprint{CTP-SCU/2018006}
\title{Minimal Length Effects on Chaotic Motion of Particles around Black Hole Horizon}
\author{Fenghua Lu}
\email{2013322020001@stu.scu.edu.cn}
\author{Jun Tao}
\email{taojun@scu.edu.cn}
\author{Peng Wang}
\email{pengw@scu.edu.cn}
\affiliation{Center for Theoretical Physics, College of Physical Science and Technology,
Sichuan University, Chengdu, 610064, PR China}

\begin{abstract}
Recently, it was conjectured that the Lyapunov exponent of chaotic motion of a
particle in a black hole is universally bounded from above by the surface
gravity of the black hole. On the other hand, the minimal length appears in
various theories of quantum gravity and leads to the deformed canonical
position-momentum commutation relation. In this paper, we use the
Hamilton-Jacobi method to study effects of the minimal length on the motion of
a massive particle perturbed away from an unstable equilibrium near the black
hole horizon. We find that the minimal length effects make the particle move
faster away from the equilibrium, and hence the corresponding Lyapunov
exponent is greater than that in the usual case with the absence of the
minimal length. It therefore shows that if the minimal length effects are
taken into account, the conjectured universal bound on the Lyapunov exponent
could be violated.

\end{abstract}
\maketitle
\tableofcontents

\bigskip

%\affiliation{Center for Theoretical Physics, College of Physical Science and Technology,
%Sichuan University, Chengdu, 610064, PR China}

%\affiliation{Center for Theoretical Physics, College of Physical Science and Technology,
%Sichuan University, Chengdu, 610064, PR China}

\section{Introduction}

Various quantum theories of gravity such as string theory
\cite{IN-Veneziano:1986zf,IN-Gross:1987ar,IN-Amati:1988tn,IN-Garay:1994en}
predict the existence of a minimal measurable length. To incorporate the
minimal length into quantum mechanics, the generalized uncertainty principle
(GUP) \cite{IN-Maggiore:1993kv,IN-Kempf:1994su} has been proposed, in which
there is a non-zero lower bound for the uncertainty in position. Moreover, the
GUP can lead to the minimal length deformed fundamental commutation relation.
For a $1$D quantum system, the deformed commutator between position and
momentum can assume the following form
\begin{equation}
\lbrack X,P]=i\hbar(1+\beta P^{2}), \label{eq:1dGUP}%
\end{equation}
where $\beta$ is some deformation parameter, and the minimal measurable length
is $\Delta X_{\min}=\hbar\sqrt{\beta}$. The deformed quantum mechanics with
modification of the usual canonical commutation relations has been
investigated intensively for various quantum systems, e.g. the harmonic
oscillator \cite{IN-Chang:2001kn}, Coulomb potential
\cite{IN-Akhoury:2003kc,IN-Brau:1999uv}, and gravitational well
\cite{IN-Brau:2006ca,IN-Pedram:2011xj}, quantum optics
\cite{IN-Pikovski:2011zk,IN-Bosso:2018ckz} and compact stars
\cite{IN-Wang:2010ct,IN-Ong:2018zqn}.

In the classical limit $\hbar\rightarrow0$, the deformed quantum mechanical
commutator is replaced by the deformed\ Poisson bracket for corresponding
classical variables:%
\begin{equation}
\frac{1}{i\hbar}\left[  \hat{A},\hat{B}\right]  \Rightarrow\left\{
A,B\right\}  ,
\end{equation}
via which effects of the minimal length can be investigated in the classical
context. The effects of the minimal length on the observational tests of
general relativity, which have been performed on Earth and in the solar
system, have been considered in
\cite{IN-Benczik:2002tt,IN-Ahmadi:2014cga,IN-Silagadze:2009vu,IN-Scardigli:2014qka,IN-Ali:2015zua,IN-Khodadi:2017eim,IN-Scardigli:2018qce}%
. The minimal length effects on the classical system were also discussed for
quantum cosmology \cite{IN-Jalalzadeh:2014jea,IN-Djordjevic:2018mzs},
classical harmonic oscillator \cite{IN-Quintela:2015bua}, and equivalence
principle \cite{IN-Tkachuk:2013qa}, Newtonian potential
\cite{IN-Scardigli:2016pjs} and the Schroinger-Newton equation
\cite{IN-Zhao:2017xjj}.

Alternatively, the classical limit of deformed Newtonian dynamics and general
relativity can be investigated using the Hamilton-Jacobi method. The deformed
Hamilton-Jacobi equation is obtained from the WKB limit of the deformed
quantum mechanics. In \cite{IN-Tao:2012fp}, we discussed the Hamilton-Jacobi
method in the context of deformed $1$D Newtonian mechanics. Later, we used the
Hamilton-Jacobi method to investigate effects of the minimal length on the
classical orbits of particles in a gravitation field, i.e. the precession of
planetary orbits in the context of deformed Newtonian dynamics and the
precession angle of planetary orbits, deflection angle of light, and time
delay in radar propagation in the context of deformed general relativity
\cite{IN-Guo:2015ldd}. Our result for the precession of planetary orbits in
the context of deformed Newtonian dynamics agreed with that in
\cite{IN-Benczik:2002tt}, in which the deformed Poisson bracket was used.
Moreover, the minimal length corrections to the Hawking temperature were also
studied using the Hamilton-Jacobi method in
\cite{IN-Chen:2013tha,IN-Chen:2013ssa,IN-Chen:2014xgj}.

Chaos in general relativity is an intriguing and important topic. Chaos is
often used to study various nonlinear phenomena in nature, and one of chaotic
behavior is that dynamical systems are highly sensitive to initial conditions.
Examples of chaotic behavior of geodesic motion in various backgrounds were
considered in
\cite{IN-Sota:1995ms,IN-Hanan:2006uf,IN-Gair:2007kr,IN-Witzany:2015yqa,IN-Wang:2016wcj,IN-Chen:2016tmr,IN-Liu:2018bmn}%
. Instead of point particles, the chaotic behavior of the ring string was
studied in \cite{IN-Frolov:1999pj,IN-Zayas:2010fs,IN-Ma:2014aha}. Recently,
the motion of a particle near the horizon of the most general static black
hole has been studied in \cite{IN-Hashimoto:2016dfz,IN-Dalui:2018qqv}, in
which the Lyapunov exponent for the motion restricted to the radial direction
was found to be the surface gravity $\kappa$ of the black hole. It was further
argued that there is a universal bound for the Lyapunov exponent of chaotic
motions of particles in black holes:%
\begin{equation}
\lambda\leq\kappa\text{.} \label{eq:LBound}%
\end{equation}
Interestingly, this bound agrees with the bound predicted in
\cite{IN-Maldacena:2015waa} for quantum field theories. However, it showed in
\cite{IN-Zhao:2018wkl} that the bound $\left(  \ref{eq:LBound}\right)  $ can
be violated for a charged massive particle perturbed from an unstable
equilibrium in some charged black hole when the equilibrium is not in the
near-horizon region.

In this paper, we study quantum gravity effects on the bound $\left(
\ref{eq:LBound}\right)  $ for motions of particles in a static black hole.
Specifically, we use the deformed Hamilton-Jacobi equation to calculate the
minimal length effects on the motion along radial direction of a massive
particle perturbed away from an unstable equilibrium near the black hole
horizon. We find that the corresponding Lyapunov exponent receives positive
correction due to the minimal length, which violates the bound $\left(
\ref{eq:LBound}\right)  $. For simplicity, we set $\hbar=c=k_{B}=1$ in this paper.

\section{Rolling Solutions of Particles near Black Hole Horizon}

\label{Sec:RSPBHH}

In this section, we use Hamilton-Jacobi method to study the motion of a
particle of mass $m$ in the near-horizon region of black holes. To be generic,
we will consider a $4$D spherically symmetric background metric of the form%
\begin{equation}
ds^{2}=-h\left(  r\right)  dt^{2}+\frac{dr^{2}}{g\left(  r\right)  }%
+r^{2}\left(  d\theta^{2}+\sin^{2}\theta d\phi^{2}\right)  , \label{eq:metric}%
\end{equation}
where $h\left(  r\right)  $ and $g\left(  r\right)  $ are assumed to have a
simple zero at the event horizon $r=r_{+}$. At $r=r_{+}$, $h\left(  r\right)
$ and $g\left(  r\right)  $ can be Taylor expanded as
\begin{align}
g\left(  r\right)   &  =\gamma\left(  r-r_{+}\right)  +\cdots,\nonumber\\
h\left(  r\right)   &  =\eta\left(  r-r_{+}\right)  +\cdots. \label{eq:NHgh}%
\end{align}
In this case, the surface gravity is given by%
\begin{equation}
\kappa=\frac{\sqrt{\gamma\eta}}{2}.
\end{equation}
As in \cite{IN-Hashimoto:2016dfz}, an external potential $V\left(  r\right)  $
is introduced such that there is an unstable equilibrium outside the horizon
for the particle. The presence of the unstable equilibrium means that any
perturbation of the particle away from the equilibrium position causes it to
roll down the effective potential. It was conjectured in
\cite{IN-Hashimoto:2016dfz} that such rolling solutions could put a universal
bound on the Lyapunov exponent of chaotic motions around black holes. We
assume that the potential is regular at the horizon and hence can set the zero
of the potential at the horizon. So at $r=r_{+}$, $V\left(  r\right)  $ is
expanded as%
\begin{equation}
V\left(  r\right)  =-v\left(  r-r_{+}\right)  +\cdots, \label{eq:NHV}%
\end{equation}
where $v$ is some positive number since one needs a repulsive force to prevent
the particle from falling into the black hole. As noted in
\cite{IN-Hashimoto:2016dfz}, the external force associated with $V\left(
r\right)  $ could be electromagnetic or scalar force.

In the remainder of this section, we calculate the particle's rolling solution
away from the unstable equilibrium near the horizon. We begin by computing the
rolling solution in the usual case. Although such solution has been obtained
in \cite{IN-Hashimoto:2016dfz}, we here focus on applying the Hamilton-Jacobi
method to find the solution. Then in the context of the minimal length
deformed general relativity, the rolling solution are investigated via the
Hamilton-Jacobi method.

\subsection{Usual Case}

\label{Sec:uc}

When a relativistic particle of mass $m$ is moving under a central potential
$V\left(  r\right)  $ in the metric $\left(  \ref{eq:metric}\right)  $, the
corresponding Hamilton-Jacobi equation is
\begin{equation}
-\frac{1}{h\left(  r\right)  }\left[  \frac{\partial S}{\partial t}+V\left(
r\right)  \right]  ^{2}+g\left(  r\right)  \left(  \partial_{r}S\right)
^{2}+\frac{\left(  \partial_{\theta}S\right)  ^{2}}{r^{2}}+\frac{\left(
\partial_{\phi}S\right)  ^{2}}{r^{2}\sin^{2}\theta}+m^{2}=0,
\end{equation}
where $S$ is the classical action. Since there are no explicit $t$-dependence
in the Hamilton-Jacobi equation, we assume that%
\begin{equation}
S=-Et+W\left(  r\right)  +\Theta\left(  \theta,\phi\right)  ,
\end{equation}
where $E$ has the meaning of the energy. To separate the variable $\theta$ and
$\phi$ from $r$, one can introduce a constant $L$ and has the equation for
$\Theta\left(  \theta,\phi\right)  $
\begin{equation}
\left(  \frac{\partial\Theta}{\partial\theta}\right)  ^{2}+\frac{\left(
\frac{\partial\Theta}{\partial\phi}\right)  ^{2}}{\sin^{2}\theta}=L^{2},
\end{equation}
where $L$ represents the orbital angular momentum. For simplicity, we only
consider the $L=0$ case, in which the particle moves in the radial direction.
The equation for $W\left(  r\right)  $ then becomes%
\begin{equation}
\frac{E-V\left(  x\right)  }{\sqrt{h\left(  r\right)  }}=\sqrt{g\left(
r\right)  p_{r}^{2}+m^{2}}, \label{eq:UcRE}%
\end{equation}
where we define $p_{r}=W^{\prime}\left(  r\right)  $. For the motion of the
particle around the equilibrium with $p_{r}^{2}\ll m^{2}$, we can focus on the
non-relativistic limit. In this limit, eqn. $\left(  \ref{eq:UcRE}\right)  $
reduces to
\begin{equation}
\frac{g\left(  r\right)  \sqrt{h\left(  r\right)  }p_{r}^{2}}{2m}%
+V_{\text{eff}}\left(  r\right)  =E, \label{eq:ucE}%
\end{equation}
where one defines the effective potential for later use:
\begin{equation}
V_{\text{eff}}\left(  r\right)  =V\left(  r\right)  +m\sqrt{h\left(  r\right)
}.
\end{equation}

The time-dependence of the motion is then obtained by the inverse Legendre
transformation:%
\begin{equation}
t=\frac{\partial W\left(  r\right)  }{\partial E},
\end{equation}
which leads to%
\begin{equation}
\dot{r}\equiv\frac{dr}{dt}=\left(  \frac{\partial p_{r}}{\partial E}\right)
^{-1}. \label{eq:tdE}%
\end{equation}
Solving eqn. $\left(  \ref{eq:ucE}\right)  $ for $p_{r}$ in terms of $E$ and
plugging it into eqn. $\left(  \ref{eq:tdE}\right)  $, we find%
\begin{equation}
\frac{m\dot{r}^{2}}{2g\left(  r\right)  \sqrt{h\left(  r\right)  }%
}+V_{\text{eff}}\left(  r\right)  =E. \label{eq:UcNE}%
\end{equation}
At the equilibrium $r=r_{0}$, one has that $E=V_{\text{eff}}\left(
r_{0}\right)  $ and $V_{\text{eff}}^{\prime}\left(  r_{0}\right)  =0$. Then
around $r=r_{0}$, eqn. $\left(  \ref{eq:UcNE}\right)  $ can be expanded as%
\begin{equation}
\dot{\epsilon}^{2}\approx-\frac{V_{\text{eff}}^{\prime\prime}\left(
r_{0}\right)  g\left(  r_{0}\right)  \sqrt{h\left(  r_{0}\right)  }%
\epsilon^{2}}{m}, \label{eq:UCepsilon}%
\end{equation}
where $\epsilon=r-r_{0}$. Near the horizon, one can use eqns. $\left(
\ref{eq:NHgh}\right)  $ and $\left(  \ref{eq:NHV}\right)  $ to show that the
effective potential $V_{\text{eff}}\left(  r\right)  $ has an unstable
equilibrium at $r_{0}=r_{+}+\frac{\eta m^{2}}{4v^{2}}$. In this case, eqn.
$\left(  \ref{eq:UCepsilon}\right)  $ then becomes%
\begin{equation}
\dot{\epsilon}=\pm\kappa\epsilon,
\end{equation}
where the plus sign corresponds to the rolling solution since the further the
particle departs from the unstable equilibrium, the faster it moves away from
the equilibrium. So the rolling solution is
\begin{equation}
\epsilon\left(  t\right)  =Ae^{\kappa t}, \label{eq:Ucrs}%
\end{equation}
where $A$ is the constant of integration.

To find the Lyapunov exponent of the rolling solution $\left(  \ref{eq:Ucrs}%
\right)  $, we consider the corresponding Jacobian matrix $K$ (it is a number
in our case), which describes the evolution of the tangent vectors:
\begin{equation}
K=\kappa.
\end{equation}
To describes how a small change of $\epsilon\left(  0\right)  $ propagates to
$\epsilon\left(  t\right)  $, we define $M\left(  t\right)  =\partial
\epsilon\left(  t\right)  /\partial\epsilon\left(  0\right)  $, which
satisfies%
\begin{equation}
\frac{dM\left(  t\right)  }{dt}=KM\left(  t\right)  .
\end{equation}
Since $M\left(  0\right)  =1$, the solution of the above equation is%
\begin{equation}
M\left(  t\right)  =e^{\kappa t}.
\end{equation}
The Lyapunov exponent $\lambda$ can calculated from $M\left(  t\right)  $:
\begin{equation}
\lambda=\lim_{t\rightarrow\infty}\frac{\ln M\left(  t\right)  }{t}=\kappa.
\end{equation}
Note that $\lambda>0$, which means that the difference between closely spaced
initial conditions grows with evolution and hence is a signature of chaos.

\subsection{Minimal Length Deformed Case}

In three dimensions, a generalization of the deformed algebra $\left(
\ref{eq:1dGUP}\right)  $ reads \cite{IN-Kempf:1994su}
\begin{align}
\lbrack X_{i},P_{j}]  &  =i\hbar\left[  (1+\beta P^{2})\delta_{ij}%
+\beta^{\prime}P_{i}P_{j}\right]  \text{,}\nonumber\\
\left[  X_{i},X_{j}\right]   &  =i\hbar\frac{\left(  2\beta-\beta^{\prime
}\right)  +\left(  2\beta+\beta^{\prime}\right)  \beta P^{2}}{1+\beta P^{2}%
}\left(  P_{i}X_{j}-P_{j}X_{i}\right)  \text{,}\\
\left[  P_{i},P_{j}\right]   &  =0\text{,}\nonumber
\end{align}
where $\beta$, $\beta^{\prime}>0$ are two deformation parameters, and the
minimal length becomes $\Delta X_{\min}=\hbar\sqrt{\beta+\beta^{\prime}}$. In
this paper, we consider the Brau reduction \cite{IN-Brau:1999uv}, where
$\beta^{\prime}=2\beta$ and the commutators taken between different components
of the position $X_{i}$ vanish to the first order in $\beta$ and
$\beta^{\prime}$. For this particular case, there is a very simple reduction
of the form to the first order in $\beta$:%
\begin{equation}
X_{i}=x_{i},\text{ }P_{i}=p_{i}\left(  1+\beta p^{2}\right)  ,
\end{equation}
where $x_{i}$ and $p_{i}$ are the conventional momentum and position operators
satisfying
\begin{equation}
\left[  x_{i},p_{j}\right]  =i\hbar\delta_{ij}\text{, }\left[  x_{i}%
,x_{j}\right]  =\left[  p_{i},p_{j}\right]  =0\text{,}%
\end{equation}
and $p^{2}=%
%TCIMACRO{\dsum \limits_{i}}%
%BeginExpansion
{\displaystyle\sum\limits_{i}}
%EndExpansion
p_{i}p_{i}$.

In \cite{IN-Guo:2015ldd}, the deformed Hamilton-Jacobi equation for
relativistic particles was derived by calculating the WKB limit of the
corresponding deformed Klein-Gordon, Dirac and Maxwell's equations. Taking the
potential $V\left(  r\right)  $ into account, we find that the deformed
relativistic Hamilton-Jacobi equation to $\mathcal{O}\left(  \beta\right)  $
in the metric $\left(  \ref{eq:metric}\right)  $ is given by
\begin{equation}
\frac{1}{h\left(  r\right)  }\left[  \frac{\partial S}{\partial t}+V\left(
r\right)  \right]  ^{2}-\mathcal{X}\left(  1+2\beta\mathcal{X}\right)  =m^{2},
\end{equation}
where we define
\begin{equation}
\mathcal{X}=g\left(  r\right)  \left(  \partial_{r}S\right)  ^{2}%
+\frac{\left(  \partial_{\theta}S\right)  ^{2}}{r^{2}}+\frac{\left(
\partial_{\phi}S\right)  ^{2}}{r^{2}\sin^{2}\theta}.
\end{equation}
As in the usual case, we only consider the case with the particle moving along
the radial direction. Therefore, separation of variables is done in the
following simple manner:%
\begin{equation}
S=-Et+W\left(  r\right)  \text{,}%
\end{equation}
where $E$ is the energy of the particle. In the non-relativistic limit,
solving the deformed Hamilton-Jacobi equation to $\mathcal{O}\left(
\beta\right)  $ gives%

\begin{equation}
p_{r}^{2}\equiv W^{\prime}\left(  r\right)  =\frac{2m}{g\left(  r\right)
}\left(  \frac{E-V_{\text{eff}}\left(  r\right)  }{\sqrt{h\left(  r\right)  }%
}\right)  \left[  1-4\beta m\left(  \frac{E-V_{\text{eff}}\left(  r\right)
}{\sqrt{h\left(  r\right)  }}\right)  \right]  .
\end{equation}
The time derivative of $r$ can be obtained by eqn. $\left(  \ref{eq:tdE}%
\right)  $:%
\begin{equation}
\frac{m^{2}\dot{r}^{2}}{g\left(  r\right)  h\left(  r\right)  }=2m\left(
\frac{E-V_{\text{eff}}\left(  r\right)  }{\sqrt{h\left(  r\right)  }}\right)
\left[  1+36\beta m\left(  \frac{E-V_{\text{eff}}\left(  r\right)  }%
{\sqrt{h\left(  r\right)  }}\right)  \right]  .
\end{equation}
Expanding the above equation around the equilibrium of $V_{\text{eff}}\left(
r\right)  $, $r=r_{0}$, one has%
\begin{equation}
-\frac{V_{\text{eff}}^{\prime\prime}\left(  r_{0}\right)  g\left(
r_{0}\right)  \sqrt{h\left(  r_{0}\right)  }}{m}\epsilon^{2}\approx
\dot{\epsilon}^{2}\left(  1-\sigma\dot{\epsilon}^{2}\right)  ,
\label{eq:gupepsilon}%
\end{equation}
where we define%
\begin{equation}
\sigma=\frac{6\beta m^{2}}{g\left(  r_{0}\right)  h\left(  r_{0}\right)  }.
\end{equation}
Near the horizon, to $\mathcal{O}\left(  \beta\right)  $, eqn. $\left(
\ref{eq:gupepsilon}\right)  $ reduces to%
\begin{equation}
\dot{\epsilon}=\pm\kappa\epsilon\left(  1+\frac{\sigma}{2}\kappa^{2}%
\epsilon^{2}\right)  ,
\end{equation}
where $\sigma=\frac{48\beta v^{4}}{\kappa^{2}\eta^{2}m^{2}}$ in this case. We
pick up the plus sign for the rolling solution, and the solution is%
\begin{equation}
\epsilon\left(  t\right)  =\epsilon_{0}\left(  t\right)  \left[
1+\frac{\sigma\kappa^{2}}{2}\epsilon_{0}\left(  t\right)  \right]  ,
\end{equation}
where $\epsilon_{0}\left(  t\right)  =Ae^{\kappa t}$ and, higher order terms
are discarded. Since $\sigma>0$, the effects of the minimal length tend to
make the particle move faster away from the equilibrium.

The Jacobian matrix for the rolling solution is%
\begin{equation}
K=\kappa\left[  1+\frac{3\sigma}{2}\kappa^{2}\epsilon^{2}\left(  t\right)
\right]  .
\end{equation}
So the corresponding $M\left(  t\right)  $ is given by%
\begin{equation}
M\left(  t\right)  =\exp\left[  \kappa t+\frac{3}{4}\sigma\kappa^{2}%
A^{2}\left(  e^{2\kappa t}-1\right)  \right]  .
\end{equation}
To find the Lyapunov exponent, we consider the finite-time Lyapunov exponent
$\lambda\left(  t\right)  $:
\begin{equation}
\lambda\left(  t\right)  \equiv\frac{\ln M\left(  t\right)  }{t}=\kappa\left[
1+\frac{3\sigma\kappa A^{2}\left(  e^{2\kappa t}-1\right)  }{4t}\right]
>\kappa. \label{eq:gupM}%
\end{equation}
If we naively take the limit of $\lambda\left(  t\right)  $ as $t$ approaches
infinity, we find the Lyapunov exponent becomes positive infinity. However
long before $t$ reach infinity, our effective approach has already broken down
when contributions from higher order terms become important. Therefore, one
might need to resort to full theory to calculate the Lyapunov exponent. On the
other hand, we can use eqn. $\left(  \ref{eq:gupM}\right)  $ to estimate the
lower bound of the Lyapunov exponent. Actually noting that $\left(  e^{2\kappa
t}-1\right)  /t$ has a minimum value of $2\kappa$ at $t=0$, one obtains that,
to $\mathcal{O}\left(  \beta\right)  $, the Lyapunov exponent $\lambda$ is
bounded from below as
\begin{equation}
\lambda\gtrsim\kappa\left(  1+\frac{3\sigma\kappa^{2}A^{2}}{2}\right)  .
\label{eq:GUPlamda}%
\end{equation}
We can further estimate $A$ by assuming that the perturbation of the particle
at the equilibrium is due to thermal fluctuations of Hawking radiation. In
this case, one finds that%
\begin{equation}
\frac{1}{2}m\dot{\epsilon}^{2}\sim\frac{1}{2}\frac{\kappa}{2\pi\sqrt{\eta
}\left(  r-r_{0}\right)  ^{1/2}}\Rightarrow A^{2}\sim\frac{v}{\pi m^{2}%
\kappa\eta}, \label{eq:GUPA}%
\end{equation}
where the equipartition theorem is used to estimate the kinetic energy of the
particle at $r=r_{0}$. Finally, eqns. $\left(  \ref{eq:GUPlamda}\right)  $ and
$\left(  \ref{eq:GUPA}\right)  $ gives that
\begin{equation}
\lambda\gtrsim\kappa\left(  1+\frac{72\beta v^{5}}{\pi\kappa\eta^{3}m^{4}%
}\right)  +\mathcal{O}\left(  \beta^{2}\right)  . \label{eq:GUPL}%
\end{equation}

\section{Discussion and Conclusion}

\label{Sec:Con}

In this paper, we have used the Hamilton-Jacobi method to investigate effects
of the minimal length on the rolling solution of a massive particle near the
horizon of a spherically symmetric black hole. An external potential was
introduced to put the particle at the unstable equilibrium outside the
horizon, and the rolling solution describes the particle's departure from the
equilibrium. In the framework of general relativity with the absence of the
minimal length, the Lyapunov exponent $\lambda$ is just the surface gravity
$\kappa$ of the black hole for such rolling solution. In the context of
deformed general relativity, we found that the presence of the minimal length
accelerates the rolling process, and the corresponding Lyapunov exponent
$\lambda>\kappa$. Our results suggest that quantum effects could make the
classical trajectory in black holes more chaotic.

In \cite{IN-Maldacena:2015waa}, it was conjectured that the Lyapunov exponent
$\lambda$ of out-of-time-ordered correlators in thermal quantum systems with a
large number of degrees of freedom is bounded by%
\begin{equation}
\lambda\leq2\pi T, \label{eq:lamdaB}%
\end{equation}
where $T=\kappa/2\pi$ is the Hawking temperature. This bound is compatible
with the universal bound for the Lyapunov exponent of chaotic motions of a
single particle in black holes, which was conjectured in
\cite{IN-Hashimoto:2016dfz,IN-Dalui:2018qqv}. In this paper, we showed that
the bound $\left(  \ref{eq:lamdaB}\right)  $ is violated for a particle moving
near the black hole horizon if effects of the minimal length are considered.
On the other hand, the presence of the minimal length could also modify the
Hawking temperature of the black hole. In fact, it showed in
\cite{IN-Chen:2014xgj} that the modified Hawking temperature is given by
\begin{equation}
T=\left[  1-\frac{\beta\left(  m^{2}+4\omega^{2}\right)  }{4}\right]  T_{0},
\end{equation}
where $T_{0}$ is the original Hawking temperature, and $m$ and $\omega$ are
the mass and the energy of radiated particles, respectively. Said differently,
effects of the minimal length decrease the Hawking temperature, which makes
the violation of the bound $\left(  \ref{eq:lamdaB}\right)  $ even worse.

In this paper, we investigated motion of a single particle outside the
horizon, and hence our results do not necessarily mean the bound $\left(
\ref{eq:lamdaB}\right)  $ conjectured in \cite{IN-Maldacena:2015waa} is
violated. However in a holographic framework, it is tempting to find how the
corrections due to the minimal length in bulk changes the bound $\left(
\ref{eq:lamdaB}\right)  $. Note that higher derivative corrections due to the
minimal length for a scalar field theory in AdS space was analyzed and related
to the conformal field theory on the boundary in \cite{CON-Faizal:2017dlb}.

\begin{acknowledgments}
We are grateful to Houwen Wu and Zheng Sun for useful discussions. This work
is supported in part by NSFC (Grant No. 11005016, 11175039, 11875196 and 11375121).
\end{acknowledgments}

\end{document}